\documentclass[10pt,conference,letterpaper,nofonttune]{IEEEtran}
\IEEEoverridecommandlockouts
\usepackage{cite}
\usepackage{amsmath,amssymb,amsfonts}
\usepackage{algorithmic}
\usepackage{graphicx}
\usepackage{textcomp}
\usepackage{xcolor}
\usepackage{xspace}
\usepackage[font=footnotesize]{caption}
\usepackage[]{subfig}
\usepackage{tabularx}
\usepackage{booktabs}
\usepackage{siunitx}
\usepackage{pgfplots}
\usepackage{pgfplotstable, xstring}
\usepackage{url}
\usetikzlibrary{pgfplots.statistics}

\newcommand{\bnset}{\ensuremath{\mathcal{N}^*}\xspace}
\newcommand{\beset}{\ensuremath{\mathcal{E}^*}\xspace}
\newcommand{\bgraph}{\ensuremath{G^*(\vnset,\beset)}\xspace}
\newcommand{\vnset}{\ensuremath{\mathcal{N}_v}\xspace}
\newcommand{\veset}{\ensuremath{\mathcal{E}_v}\xspace}

\newcommand{\vgraph}{\ensuremath{G(\vnset,\veset)}\xspace}

\newcommand{\gnb}{\gls{gnb}\xspace}
\newcommand{\gnbs}{\glspl{gnb}\xspace}

\newcommand{\uilk}{\ensuremath{u_{i,l,k}}\xspace}

\newcommand{\pijk}{\ensuremath{P_{i,j,k}}\xspace}
\newcommand{\eij}{\ensuremath{e_{i,j}}\xspace}
\newcommand{\D}{\ensuremath{D}\xspace}
\newcommand{\dg}{\ensuremath{\delta}\xspace}

\newcommand{\los}{LoS\xspace}

\newcommand{\red}{\ensuremath{R}\xspace}
\newcommand{\capi}{\ensuremath{C}\xspace}
\newcommand{\demi}{\ensuremath{d_i}\xspace}
\newcommand{\capij}{\ensuremath{L_{i,j}}\xspace}
\newcommand{\fijh}{\ensuremath{f_{i,j,h}}\xspace}
\newcommand{\fijhk}{\ensuremath{f_{i,j,h,k}}\xspace}
\newcommand{\luij}{\ensuremath{a_{i,j}}\xspace}

\newcommand{\ran}{\gls{ran}\xspace}

\newcommand{\nonrt}{\gls{non-rt}\xspace}

\newcommand{\ric}{\gls{ric}\xspace}

\newcommand{\iabd}{\gls{iab}-donor\xspace}
\newcommand{\iabds}{\gls{iab}-donors\xspace}
\newcommand{\iabn}{\gls{iab}-node\xspace}
\newcommand{\iabns}{\gls{iab}-nodes\xspace}

\newcommand{\rev}[1]{\textcolor{black}{#1}}

\usepackage[acronyms,nonumberlist,nopostdot,nomain,nogroupskip,acronymlists={hidden}]{glossaries}
\newglossary[algh]{hidden}{acrh}{acnh}{Hidden Acronyms}

\newacronym{3gpp}{3GPP}{3rd Generation Partnership Project}
\newacronym{4g}{4G}{4th generation}
\newacronym{5g}{5G}{5th generation}
\newacronym{6g}{6G}{6th generation}
\newacronym{5gc}{5GC}{5G Core}
\newacronym{adc}{ADC}{Analog to Digital Converter}
\newacronym{aerpaw}{AERPAW}{Aerial Experimentation and Research Platform for Advanced Wireless}
\newacronym{ai}{AI}{Artificial Intelligence}
\newacronym{aimd}{AIMD}{Additive Increase Multiplicative Decrease}
\newacronym{am}{AM}{Acknowledged Mode}
\newacronym{amc}{AMC}{Adaptive Modulation and Coding}
\newacronym{amf}{AMF}{Access and Mobility Management Function}
\newacronym{aops}{AOPS}{Adaptive Order Prediction Scheduling}
\newacronym{api}{API}{Application Programming Interface}
\newacronym{apn}{APN}{Access Point Name}
\newacronym{ap}{AP}{Application Protocol}
\newacronym{aqm}{AQM}{Active Queue Management}
\newacronym{ar}{AR}{Augmented Reality}
\newacronym{ausf}{AUSF}{Authentication Server Function}
\newacronym{avc}{AVC}{Advanced Video Coding}
\newacronym{awgn}{AGWN}{Additive White Gaussian Noise}
\newacronym{balia}{BALIA}{Balanced Link Adaptation Algorithm}
\newacronym{bbu}{BBU}{Base Band Unit}
\newacronym{bdp}{BDP}{Bandwidth-Delay Product}
\newacronym{ber}{BER}{Bit Error Rate}
\newacronym{bf}{BF}{Beamforming}
\newacronym{bler}{BLER}{Block Error Rate}
\newacronym{brr}{BRR}{Bayesian Ridge Regressor}
\newacronym{bs}{BS}{Base Station}
\newacronym{bsr}{BSR}{Buffer Status Report}
\newacronym{bss}{BSS}{Business Support System}
\newacronym{ca}{CA}{Carrier Aggregation}
\newacronym{caas}{CaaS}{Connectivity-as-a-Service}
\newacronym{cb}{CB}{Code Block}
\newacronym{cc}{CC}{Congestion Control}
\newacronym{ccid}{CCID}{Congestion Control ID}
\newacronym{cco}{CC}{Carrier Component}
\newacronym{cdd}{CDD}{Cyclic Delay Diversity}
\newacronym{cdf}{CDF}{Cumulative Distribution Function}
\newacronym{cdn}{CDN}{Content Distribution Network}
\newacronym{cn}{CN}{Core Network}
\newacronym{codel}{CoDel}{Controlled Delay Management}
\newacronym{comac}{COMAC}{Converged Multi-Access and Core}
\newacronym{cord}{CORD}{Central Office Re-architected as a Datacenter}
\newacronym{cornet}{CORNET}{COgnitive Radio NETwork}
\newacronym{cosmos}{COSMOS}{Cloud Enhanced Open Software Defined Mobile Wireless Testbed for City-Scale Deployment}
\newacronym{cots}{COTS}{Commercial Off-the-Shelf}
\newacronym{cp}{CP}{Control Plane}
\newacronym{cyp}{CP}{Cyclic Prefix}
\newacronym{up}{UP}{User Plane}
\newacronym{cpu}{CPU}{Central Processing Unit}
\newacronym{cqi}{CQI}{Channel Quality Information}
\newacronym{cr}{CR}{Cognitive Radio}
\newacronym{cran}{C-RAN}{Cloud \gls{ran}}
\newacronym{crs}{CRS}{Cell Reference Signal}
\newacronym{csi}{CSI}{Channel State Information}
\newacronym{csirs}{CSI-RS}{Channel State Information - Reference Signal}
\newacronym{cu}{CU}{Central Unit}
\newacronym{d2tcp}{D$^2$TCP}{Deadline-aware Data center TCP}
\newacronym{d3}{D$^3$}{Deadline-Driven Delivery}
\newacronym{dac}{DAC}{Digital to Analog Converter}
\newacronym{dag}{DAG}{Directed Acyclic Graph}
\newacronym{das}{DAS}{Distributed Antenna System}
\newacronym{dash}{DASH}{Dynamic Adaptive Streaming over HTTP}
\newacronym{dc}{DC}{Dual Connectivity}
\newacronym{dccp}{DCCP}{Datagram Congestion Control Protocol}
\newacronym{dce}{DCE}{Direct Code Execution}
\newacronym{dci}{DCI}{Downlink Control Information}
\newacronym{dctcp}{DCTCP}{Data Center TCP}
\newacronym{dl}{DL}{Downlink}
\newacronym{dmr}{DMR}{Deadline Miss Ratio}
\newacronym{dmrs}{DMRS}{DeModulation Reference Signal}
\newacronym{drlcc}{DRL-CC}{Deep Reinforcement Learning Congestion Control}
\newacronym{drs}{DRS}{Discovery Reference Signal}
\newacronym{du}{DU}{Distributed Unit}
\newacronym{e2e}{E2E}{end-to-end}
\newacronym{ecaas}{ECaaS}{Edge-Cloud-as-a-Service}
\newacronym{ecn}{ECN}{Explicit Congestion Notification}
\newacronym{edf}{EDF}{Earliest Deadline First}
\newacronym{embb}{eMBB}{Enhanced Mobile Broadband}
\newacronym{empower}{EMPOWER}{EMpowering transatlantic PlatfOrms for advanced WirEless Research}
\newacronym{enb}{eNB}{evolved Node Base}
\newacronym{endc}{EN-DC}{E-UTRAN-\gls{nr} \gls{dc}}
\newacronym{epc}{EPC}{Evolved Packet Core}
\newacronym{eps}{EPS}{Evolved Packet System}
\newacronym{es}{ES}{Edge Server}
\newacronym{etsi}{ETSI}{European Telecommunications Standards Institute}
\newacronym[firstplural=Estimated Times of Arrival (ETAs)]{eta}{ETA}{Estimated Time of Arrival}
\newacronym{eutran}{E-UTRAN}{Evolved Universal Terrestrial Access Network}
\newacronym{faas}{FaaS}{Function-as-a-Service}
\newacronym{fapi}{FAPI}{Functional Application Platform Interface}
\newacronym{fdd}{FDD}{Frequency Division Duplexing}
\newacronym{fdm}{FDM}{Frequency Division Multiplexing}
\newacronym{fdma}{FDMA}{Frequency Division Multiple Access}
\newacronym{fed4fire}{FED4FIRE+}{Federation 4 Future Internet Research and Experimentation Plus}
\newacronym{fir}{FIR}{Finite Impulse Response}
\newacronym{fit}{FIT}{Future \acrlong{iot}}
\newacronym{fpga}{FPGA}{Field Programmable Gate Array}
\newacronym{fr2}{FR2}{Frequency Range 2}
\newacronym{fs}{FS}{Fast Switching}
\newacronym{fscc}{FSCC}{Flow Sharing Congestion Control}
\newacronym{ftp}{FTP}{File Transfer Protocol}
\newacronym{fw}{FW}{Flow Window}
\newacronym{ge}{GE}{Gaussian Elimination}
\newacronym{gnb}{gNB}{Next Generation Node Base}
\newacronym{gop}{GOP}{Group of Pictures}
\newacronym{gpr}{GPR}{Gaussian Process Regressor}
\newacronym{gpu}{GPU}{Graphics Processing Unit}
\newacronym{gtp}{GTP}{GPRS Tunneling Protocol}
\newacronym{gtpc}{GTP-C}{GPRS Tunnelling Protocol Control Plane}
\newacronym{gtpu}{GTP-U}{GPRS Tunnelling Protocol User Plane}
\newacronym{gtpv2c}{GTPv2-C}{\gls{gtp} v2 - Control}
\newacronym{gw}{GW}{Gateway}
\newacronym{harq}{HARQ}{Hybrid Automatic Repeat reQuest}
\newacronym{hetnet}{HetNet}{Heterogeneous Network}
\newacronym{hh}{HH}{Hard Handover}
\newacronym{hol}{HOL}{Head-of-Line}
\newacronym{hqf}{HQF}{Highest-quality-first}
\newacronym{hss}{HSS}{Home Subscription Server}
\newacronym{http}{HTTP}{HyperText Transfer Protocol}
\newacronym{ia}{IA}{Initial Access}
\newacronym{iab}{IAB}{Integrated Access and Backhaul}
\newacronym{ic}{IC}{Incident Command}
\newacronym{icmp}{ICMP}{Internet Control Message Protocol}
\newacronym{ietf}{IETF}{Internet Engineering Task Force}
\newacronym{imsi}{IMSI}{International Mobile Subscriber Identity}
\newacronym{imt}{IMT}{International Mobile Telecommunication}
\newacronym{iot}{IoT}{Internet of Things}
\newacronym{ip}{IP}{Internet Protocol}
\newacronym{itu}{ITU}{International Telecommunication Union}
\newacronym{kpi}{KPI}{Key Performance Indicator}
\newacronym{kpm}{KPM}{Key Performance Measurement}
\newacronym{kvm}{KVM}{Kernel-based Virtual Machine}
\newacronym{los}{LOS}{Line-of-Sight}
\newacronym{lsm}{LSM}{Link-to-System Mapping}
\newacronym{lstm}{LSTM}{Long Short Term Memory}
\newacronym{lte}{LTE}{Long Term Evolution}
\newacronym{lxc}{LXC}{Linux Container}
\newacronym{m2m}{M2M}{Machine to Machine}
\newacronym{mac}{MAC}{Medium Access Control}
\newacronym{manet}{MANET}{Mobile Ad Hoc Network}
\newacronym{mano}{MANO}{Management and Orchestration}
\newacronym{mc}{MC}{Multi-Connectivity}
\newacronym{mcc}{MCC}{Mobile Cloud Computing}
\newacronym{mchem}{MCHEM}{Massive Channel Emulator}
\newacronym{mcs}{MCS}{Modulation and Coding Scheme}
\newacronym{mec}{MEC}{Multi-access Edge Computing}
\newacronym{mec2}{MEC}{Mobile Edge Cloud}
\newacronym{mfc}{MFC}{Mobile Fog Computing}
\newacronym{mgen}{MGEN}{Multi-Generator}
\newacronym{mi}{MI}{Mutual Information}
\newacronym{mib}{MIB}{Master Information Block}
\newacronym{miesm}{MIESM}{Mutual Information Based Effective SINR}
\newacronym{mimo}{MIMO}{Multiple Input, Multiple Output}
\newacronym{ml}{ML}{Machine Learning}
\newacronym{mlr}{MLR}{Maximum-local-rate}
\newacronym[plural=\gls{mme}s,firstplural=Mobility Management Entities (MMEs)]{mme}{MME}{Mobility Management Entity}
\newacronym{mmtc}{mMTC}{Massive Machine-Type Communications}
\newacronym{mmwave}{mmWave}{millimeter wave}
\newacronym{mpdccp}{MP-DCCP}{Multipath Datagram Congestion Control Protocol}
\newacronym{mptcp}{MPTCP}{Multipath TCP}
\newacronym{mr}{MR}{Maximum Rate}
\newacronym{mrdc}{MR-DC}{Multi \gls{rat} \gls{dc}}
\newacronym{mse}{MSE}{Mean Square Error}
\newacronym{mss}{MSS}{Maximum Segment Size}
\newacronym{mt}{MT}{Mobile Termination}
\newacronym{mtd}{MTD}{Machine-Type Device}
\newacronym{mtu}{MTU}{Maximum Transmission Unit}
\newacronym{mumimo}{MU-MIMO}{Multi-user \gls{mimo}}
\newacronym{mvno}{MVNO}{Mobile Virtual Network Operator}
\newacronym{nalu}{NALU}{Network Abstraction Layer Unit}
\newacronym{nas}{NAS}{Non-Access Stratum}
\newacronym{nbiot}{NB-IoT}{Narrow Band IoT}
\newacronym{nfv}{NFV}{Network Function Virtualization}
\newacronym{nfvi}{NFVI}{Network Function Virtualization Infrastructure}
\newacronym{ngrg}{nGRG}{next Generation Research Group}
\newacronym{ni}{NI}{Network Interfaces}
\newacronym{nic}{NIC}{Network Interface Card}
\newacronym{nlos}{NLOS}{Non-Line-of-Sight}
\newacronym{now}{NOW}{Non Overlapping Window}
\newacronym{nsm}{NSM}{Network Service Mesh}
\newacronym{nr}{NR}{New Radio}
\newacronym{nrf}{NRF}{Network Repository Function}
\newacronym{nsa}{NSA}{Non Stand Alone}
\newacronym{nse}{NSE}{Network Slicing Engine}
\newacronym{nssf}{NSSF}{Network Slice Selection Function}
\newacronym{o2i}{O2I}{Outdoor to Indoor}
\newacronym{oai}{OAI}{OpenAirInterface}
\newacronym{oaicn}{OAI-CN}{\gls{oai} \acrlong{cn}}
\newacronym{oairan}{OAI-RAN}{\acrlong{oai} \acrlong{ran}}
\newacronym{oam}{OAM}{Operations, Administration and Maintenance}
\newacronym{ofdm}{OFDM}{Orthogonal Frequency Division Multiplexing}
\newacronym{olia}{OLIA}{Opportunistic Linked Increase Algorithm}
\newacronym{omec}{OMEC}{Open Mobile Evolved Core}
\newacronym{onap}{ONAP}{Open Network Automation Platform}
\newacronym{onf}{ONF}{Open Networking Foundation}
\newacronym{onos}{ONOS}{Open Networking Operating System}
\newacronym{oom}{OOM}{\gls{onap} Operations Manager}
\newacronym{opnfv}{OPNFV}{Open Platform for \gls{nfv}}
\newacronym{oran}{O-RAN}{}
\newacronym{orbit}{ORBIT}{Open-Access Research Testbed for Next-Generation Wireless Networks}
\newacronym{os}{OS}{Operating System}
\newacronym{oss}{OSS}{Operations Support System}
\newacronym{otic}{OTIC}{Open Testing \& Integration Centre}
\newacronym{pa}{PA}{Position-aware}
\newacronym{pase}{PASE}{Prioritization, Arbitration, and Self-adjusting Endpoints}
\newacronym{pawr}{PAWR}{Platforms for Advanced Wireless Research}
\newacronym{pbch}{PBCH}{Physical Broadcast Channel}
\newacronym{pcef}{PCEF}{Policy and Charging Enforcement Function}
\newacronym{pcfich}{PCFICH}{Physical Control Format Indicator Channel}
\newacronym{pcrf}{PCRF}{Policy and Charging Rules Function}
\newacronym{pdcch}{PDCCH}{Physical Downlink Control Channel}
\newacronym{pdcp}{PDCP}{Packet Data Convergence Protocol}
\newacronym{pdsch}{PDSCH}{Physical Downlink Shared Channel}
\newacronym{pdu}{PDU}{Packet Data Unit}
\newacronym{pf}{PF}{Proportional Fair}
\newacronym{pgw}{PGW}{Packet Gateway}
\newacronym{phich}{PHICH}{Physical Hybrid ARQ Indicator Channel}
\newacronym{phy}{PHY}{Physical}
\newacronym{pmch}{PMCH}{Physical Multicast Channel}
\newacronym{pmi}{PMI}{Precoding Matrix Indicators}
\newacronym{powder}{POWDER}{Platform for Open Wireless Data-driven Experimental Research}
\newacronym{ppo}{PPO}{Proximal Policy Optimization}
\newacronym{ppp}{PPP}{Poisson Point Process}
\newacronym{prach}{PRACH}{Physical Random Access Channel}
\newacronym{prb}{PRB}{Physical Resource Block}
\newacronym{psnr}{PSNR}{Peak Signal to Noise Ratio}
\newacronym{pss}{PSS}{Primary Synchronization Signal}
\newacronym{pucch}{PUCCH}{Physical Uplink Control Channel}
\newacronym{pusch}{PUSCH}{Physical Uplink Shared Channel}
\newacronym{qam}{QAM}{Quadrature Amplitude Modulation}
\newacronym{qci}{QCI}{\gls{qos} Class Identifier}
\newacronym{qoe}{QoE}{Quality of Experience}
\newacronym{qos}{QoS}{Quality of Service}
\newacronym{quic}{QUIC}{Quick UDP Internet Connections}
\newacronym{rach}{RACH}{Random Access Channel}
\newacronym{ran}{RAN}{Radio Access Network}
\newacronym[firstplural=Radio Access Technologies (RATs)]{rat}{RAT}{Radio Access Technology}
\newacronym{rcn}{RCN}{Research Coordination Network}
\newacronym{rc}{RC}{RAN Control}
\newacronym{rec}{REC}{Radio Edge Cloud}
\newacronym{red}{RED}{Random Early Detection}
\newacronym{renew}{RENEW}{Reconfigurable Eco-system for Next-generation End-to-end Wireless}
\newacronym{rf}{RF}{Radio Frequency}
\newacronym{rfc}{RFC}{Request for Comments}
\newacronym{rfr}{RFR}{Random Forest Regressor}
\newacronym{ric}{RIC}{RAN Intelligent Controller}
\newacronym{rlc}{RLC}{Radio Link Control}
\newacronym{rlf}{RLF}{Radio Link Failure}
\newacronym{rlnc}{RLNC}{Random Linear Network Coding}
\newacronym{rmr}{RMR}{RIC Message Router}
\newacronym{rmse}{RMSE}{Root Mean Squared Error}
\newacronym{rnis}{RNIS}{Radio Network Information Service}
\newacronym{rr}{RR}{Round Robin}
\newacronym{rrc}{RRC}{Radio Resource Control}
\newacronym{rrm}{RRM}{Radio Resource Management}
\newacronym{rru}{RRU}{Remote Radio Unit}
\newacronym{rs}{RS}{Remote Server}
\newacronym{rsrp}{RSRP}{Reference Signal Received Power}
\newacronym{rsrq}{RSRQ}{Reference Signal Received Quality}
\newacronym{rss}{RSS}{Received Signal Strength}
\newacronym{rssi}{RSSI}{Received Signal Strength Indicator}
\newacronym{rtt}{RTT}{Round Trip Time}
\newacronym{ru}{RU}{Radio Unit}
\newacronym{rw}{RW}{Receive Window}
\newacronym{rx}{RX}{Receiver}
\newacronym{s1ap}{S1AP}{S1 Application Protocol}
\newacronym{sa}{SA}{standalone}
\newacronym{sack}{SACK}{Selective Acknowledgment}
\newacronym{sap}{SAP}{Service Access Point}
\newacronym{sc2}{SC2}{Spectrum Collaboration Challenge}
\newacronym{scef}{SCEF}{Service Capability Exposure Function}
\newacronym{sch}{SCH}{Secondary Cell Handover}
\newacronym{scoot}{SCOOT}{Split Cycle Offset Optimization Technique}
\newacronym{sctp}{SCTP}{Stream Control Transmission Protocol}
\newacronym{sdap}{SDAP}{Service Data Adaptation Protocol}
\newacronym{sdk}{SDK}{Software Development Kit}
\newacronym{sdm}{SDM}{Space Division Multiplexing}
\newacronym{sdma}{SDMA}{Spatial Division Multiple Access}
\newacronym{sdn}{SDN}{Software-defined Networking}
\newacronym{sdr}{SDR}{Software-defined Radio}
\newacronym{seba}{SEBA}{SDN-Enabled Broadband Access}
\newacronym{sgsn}{SGSN}{Serving GPRS Support Node}
\newacronym{sgw}{SGW}{Service Gateway}
\newacronym{si}{SI}{Study Item}
\newacronym{sib}{SIB}{Secondary Information Block}
\newacronym{sinr}{SINR}{Signal to Interference plus Noise Ratio}
\newacronym{sip}{SIP}{Session Initiation Protocol}
\newacronym{siso}{SISO}{Single Input, Single Output}
\newacronym{sla}{SLA}{Service Level Agreement}
\newacronym{sm}{SM}{Service Model}
\newacronym{smf}{SMF}{Session Management Function}
\newacronym{smo}{SMO}{Service Management and Orchestration}
\newacronym{sms}{SMS}{Short Message Service}
\newacronym{smsgmsc}{SMS-GMSC}{\gls{sms}-Gateway}
\newacronym{snr}{SNR}{Signal-to-Noise-Ratio}
\newacronym{son}{SON}{Self-Organizing Network}
\newacronym{sptcp}{SPTCP}{Single Path TCP}
\newacronym{srb}{SRB}{Service Radio Bearer}
\newacronym{srn}{SRN}{Standard Radio Node}
\newacronym{srs}{SRS}{Sounding Reference Signal}
\newacronym{ss}{SS}{Synchronization Signal}
\newacronym{sss}{SSS}{Secondary Synchronization Signal}
\newacronym{st}{ST}{Spanning Tree}
\newacronym{svc}{SVC}{Scalable Video Coding}
\newacronym{tb}{TB}{Transport Block}
\newacronym{tcp}{TCP}{Transmission Control Protocol}
\newacronym{tdd}{TDD}{Time Division Duplexing}
\newacronym{tdm}{TDM}{Time Division Multiplexing}
\newacronym{tdma}{TDMA}{Time Division Multiple Access}
\newacronym{tfl}{TfL}{Transport for London}
\newacronym{tfrc}{TFRC}{TCP-Friendly Rate Control}
\newacronym{tft}{TFT}{Traffic Flow Template}
\newacronym{tgen}{TGEN}{Traffic Generator}
\newacronym{tip}{TIP}{Telecom Infra Project}
\newacronym{tm}{TM}{Transparent Mode}
\newacronym{to}{TO}{Telco Operator}
\newacronym{tr}{TR}{Technical Report}
\newacronym{trp}{TRP}{Transmitter Receiver Pair}
\newacronym{ts}{TS}{Technical Specification}
\newacronym{tti}{TTI}{Transmission Time Interval}
\newacronym{ttt}{TTT}{Time-to-Trigger}
\newacronym{tx}{TX}{Transmitter}
\newacronym{uas}{UAS}{Unmanned Aerial System}
\newacronym{uav}{UAV}{Unmanned Aerial Vehicle}
\newacronym{udm}{UDM}{Unified Data Management}
\newacronym{udp}{UDP}{User Datagram Protocol}
\newacronym{udr}{UDR}{Unified Data Repository}
\newacronym{ue}{UE}{User Equipment}
\newacronym{uhd}{UHD}{\gls{usrp} Hardware Driver}
\newacronym{ul}{UL}{Uplink}
\newacronym{um}{UM}{Unacknowledged Mode}
\newacronym{uml}{UML}{Unified Modeling Language}
\newacronym{upa}{UPA}{Uniform Planar Array}
\newacronym{upf}{UPF}{User Plane Function}
\newacronym{urllc}{URLLC}{Ultra Reliable and Low Latency Communications}
\newacronym{usa}{U.S.}{United States}
\newacronym{usim}{USIM}{Universal Subscriber Identity Module}
\newacronym{usrp}{USRP}{Universal Software Radio Peripheral}
\newacronym{utc}{UTC}{Urban Traffic Control}
\newacronym{vim}{VIM}{Virtualization Infrastructure Manager}
\newacronym{vm}{VM}{Virtual Machine}
\newacronym{vnf}{VNF}{Virtual Network Function}
\newacronym{volte}{VoLTE}{Voice over \gls{lte}}
\newacronym{voltha}{VOLTHA}{Virtual OLT HArdware Abstraction}
\newacronym{vr}{VR}{Virtual Reality}
\newacronym{vran}{vRAN}{Virtualized \gls{ran}}
\newacronym{vss}{VSS}{Video Streaming Server}
\newacronym{wbf}{WBF}{Wired Bias Function}
\newacronym{wf}{WF}{Waterfilling}
\newacronym{wg}{WG}{Working Group}
\newacronym{wlan}{WLAN}{Wireless Local Area Network}
\newacronym{osm}{OSM}{Open Source Management and Orchestration}
\newacronym{pnf}{PNF}{Physical Network Function}
\newacronym{drl}{DRL}{Deep Reinforcement Learning}
\newacronym{mtc}{MTC}{Machine-type Communications}
\newacronym{osc}{OSC}{O-RAN Software Community}
\newacronym{mns}{MnS}{Management Services}
\newacronym{ves}{VES}{\gls{vnf} Event Stream}
\newacronym{ei}{EI}{Enrichment Information}
\newacronym{fh}{FH}{Fronthaul}
\newacronym{fft}{FFT}{Fast Fourier Transform}
\newacronym{laa}{LAA}{Licensed-Assisted Access}
\newacronym{plfs}{PLFS}{Physical Layer Frequency Signals}
\newacronym{ptp}{PTP}{Precision Time Protocol}
\newacronym{asic}{ASIC}{Application-specific Integrated Circuit}
\newacronym{aal}{AAL}{Acceleration Abstraction Layer}
\newacronym{fec}{FEC}{Forward Error Correction}
\newacronym{sdl}{SDL}{Shared Data Layer}
\newacronym{nib}{NIB}{Network Information Base}
\newacronym{rnib}{R-NIB}{RAN \gls{nib}}
\newacronym{fcaps}{FCAPS}{Fault, Configuration, Accounting, Performance, Security}
\newacronym{ie}{IE}{Information Element}
\newacronym{fg}{FG}{Focus Group}
\newacronym{osfg}{OSFG}{Open Source Focus Group}
\newacronym{sdfg}{SDFG}{Standard Development Focus Group}
\newacronym{tifg}{TIFG}{Test \& Integration Focus Group}
\newacronym{sfg}{SFG}{Security Focus Group}
\newacronym{swg}{SWG}{Security Work Group}
\newacronym{e2sm}{E2SM}{E2 Service Model}
\newacronym{tsc}{TSC}{Technical Steering Committee}
\newacronym{sdo}{SDO}{Standard-Development Organization}
\newacronym{sql}{SQL}{Structured Query Language}
\newacronym{ssh}{SSH}{Secure Shell}
\newacronym{tls}{TLS}{Transport Layer Security}
\newacronym{netconf}{NETCONF}{Network Configuration Protocol}
\newacronym{dtls}{DTLS}{Datagram Transport Layer Security}
\newacronym{cmp}{CMP}{Certificate Management Protocol}
\newacronym{ccc}{CCC}{Cell Configuration and Control}
\newacronym{dsp}{DSP}{Digital Signal Processing}
\newacronym{opex}{OPEX}{Operational Expenses}
\newacronym{cbrs}{CBRS}{Citizen Broadband Radio Service}
\newacronym{ntn}{NTN}{Non-terrestrial Network}

\newacronym{o-ran}{O-RAN}{}
\newacronym{cnn}{CNN}{Convolutional Neural Network}
\newacronym{appmgr}{APPMGR}{App Manager}
\newacronym{near-rt}{Near-RT}{Near-real-time}
\newacronym{non-rt}{Non-RT}{Non-real-time}
\newacronym{ilp}{ILP}{Integer-Linear Problem}

\usepackage[capitalise]{cleveref}

\usepackage{soul}

\begin{document}
\title{Optimizing and Managing Wireless Backhaul for Resilient Next-Generation Cellular Networks}

 \author{
 \IEEEauthorblockN{Gabriele Gemmi\IEEEauthorrefmark{1}, Michele Polese\IEEEauthorrefmark{1}, Tommaso Melodia\IEEEauthorrefmark{1}, Leonardo Maccari\IEEEauthorrefmark{2}}
 \IEEEauthorblockN{\IEEEauthorrefmark{1}Northeastern University, Boston, MA, USA, 
\IEEEauthorrefmark{2}Università Ca' Foscari, Venezia, Italy\\
Email: \{{g.gemmi, m.polese, t.melodia\}@northeastern.edu}, leonardo.maccari@unive.it
  }
  \thanks{This work was partially supported by NGIAtlantic.eu project within the EU Horizon 2020 programme under Grant No. 871582, and by OUSD(R\&E) through Army Research Laboratory Cooperative Agreement Number W911NF-19-2-0221. The views and conclusions contained in this document are those of the authors and should not be interpreted as representing the official policies, either expressed or implied, of the Army Research Laboratory or the U.S. Government. The U.S. Government is authorized to reproduce and distribute reprints for Government purposes notwithstanding any copyright notation herein.}
  }

\maketitle
\begin{abstract}

Next-generation wireless networks target high network availability, ubiquitous coverage, and extremely high data rates for mobile users. This requires exploring new frequency bands, e.g., mmWaves, moving toward ultra-dense deployments in urban locations, and providing ad hoc, resilient connectivity in rural scenarios. 
The design of the backhaul network plays a key role in advancing how the access part of the wireless system supports next-generation use cases. 
Wireless backhauling, such as the newly introduced \gls{iab} concept in 5G, provides a promising solution, also leveraging the mmWave technology and steerable beams to mitigate interference and scalability issues. 
At the same time, however, managing and optimizing a complex wireless backhaul introduces additional challenges for the operation of cellular systems.
This paper presents a strategy for the optimal creation of the backhaul network considering various constraints related to network topology, robustness, and flow management. We evaluate its feasibility and efficiency using synthetic and realistic network scenarios based on 3D modeling of buildings and ray tracing. 
We implement and prototype our solution as a dynamic \gls{iab} control framework based on the Open \gls{ran} architecture, and demonstrate its functionality in Colosseum, a large-scale wireless network emulator with hardware in the loop.

\end{abstract}

\begin{picture}(0,0)(10,-410)
    \put(0,0){
    \put(0,0){\footnotesize \scshape This paper has been accepted for publication on IEEE International Conference on Network and Service Management 2024.}
     \put(0,-10){
     \scriptsize\scshape \textcopyright~2024 IEEE. Personal use of this material is permitted. Permission from IEEE must be obtained for all other uses, in any current or future media, including}
     \put(0, -17){
     \scriptsize\scshape reprinting/republishing this material for advertising or promotional purposes, creating new collective works, for resale or redistribution to servers or}
     \put(0, -24){
     \scriptsize\scshape lists, or reuse of any copyrighted component of this work in other works.}
     }
 \end{picture}

\section{Introduction}
\label{sec:introduction}

Next-generation cellular networks will be deployed in a variety of operational environments (e.g., ultra-dense urban, rural)~\cite{yaacoub2020key}, supported by heterogeneous \glspl{rat}~\cite{saad2020vision}, and operating in different portions of the spectrum~\cite{rappaport2013millimeter}. As an example, urban cellular networks are moving toward a 10-fold increase in the density of \gnbs, from 10 \gnbs per km$^2$ typically deployed today to an estimated 100 \gnbs per km$^2$ in 5G~\cite{itu2017guidelines}. 

The diverse characteristics of these access scenarios introduce new challenges for the design of a robust and high performance backhaul network, which needs to be pervasive and flexible. Traditional backhaul solutions, based on fiber drops or point-to-point dedicated wireless links to each \gls{ran} base station, are limited in terms of scalability and cost, and are often referred to as one of the barriers toward both ultra-dense deployments in urban areas and remote, rural access. For these reasons, the \gls{3gpp} has introduced native support for wireless backhaul in Release 16 for its \gls{5g} cellular technology, i.e., \gls{3gpp} NR. Specifically, \gls{iab} introduces wireless self-backhauling with the same waveform and protocol stack already used for the access part of the network~\cite{madapatha2020integrated,polese2020integrated}. With \gls{iab}, only some of the \gnbs need to be fiber-connected (i.e., the \textit{\gls{iab}-donors}), and a wireless multi-hop network topology connects every other \gnb (i.e., the \emph{\gls{iab}-nodes})  to the closest donor. \gls{iab} networks are thus organized in multiple trees, with roots in different donors and leaves represented by the \glspl{ue}.

Wireless self-backhaul, integrated with the access network, significantly improves the flexibility of cellular deployments for 5G and beyond, but, at the same time, adds \emph{complexity in the management plane and needs to be properly designed and deployed to offer the best performance to the end users of the network}. On one hand, \gls{iab} in 5G and beyond can leverage the new spectrum above 6 GHz, in the lower \gls{mmwave} band, to increase the capacity of the system~\cite{rappaport2013millimeter} and to limit self-interference thanks to highly directional transmissions from large antenna arrays in the \gls{iab} nodes. On the other, \gls{mmwave} links present a 15 dB to 30 dB gap in received power between \gls{los} and \gls{nlos} conditions~\cite{rappaport2013millimeter,raghavan2018spatio}, thus requiring careful planning that favors \gls{los} links and avoids blockage conditions~\cite{saha2019milliter,fiore2022boosting}. Similarly, the opportunity of re-using the same frequency bands for access and backhaul provides a potential multiplexing gain, but also challenges for scheduling of data flows across the two parts of the network~\cite{gopalam2022distributed}. Finally, self-backhaul solutions extend the scenarios in which access connectivity can be provided, but at the same time a failure in a wireless backhaul link has a more significant impact than a failure of a single access link, as it impacts all the downstream nodes in the topology tree.

These reasons make design and optimization of \gls{iab} networks critical parts of next-generation wireless planning and operations. In this paper, we focus on pre-deployment and post-deployment approaches for the optimal identification and management of topologies across complex \gls{iab} networks, with the goal of (i) providing a minimum guaranteed area capacity, in line with the \gls{itu} recommendations for next-generation cellular networks~\cite{itu2017guidelines}; and (ii) minimizing the downtime of the \gls{iab} tree in case of failures of links between parent and child \gls{iab} nodes. 
Specifically, our contributions are as follows:

\begin{itemize}
    \item We introduce mixed \glspl{ilp} that combine topological, resiliency, and flow constraints, which we test on synthetic graphs and realistic topologies derived by 3D surfaces representing real world urban areas. 
    Using accessible hardware, we achieve near-optimal solutions for fault-tolerant synthetic topologies  up to 45 nodes, and up to 60 nodes for realistic topologies. In both cases we measure a lower bound on the required percentage of donors around 20\%.
    
    \item We prototype the network optimization and management routine on an O-RAN rApp (i.e., a piece of custom control logic running in the O-RAN \nonrt \gls{ric}~\cite{moro2023toward,polese2023understanding}). The rApp dynamically recreates the \gls{iab} topology in case of link failures in the backhaul topology, based on the guidance from the optimization problem. We use Colosseum~\cite{bonati2021colosseum}, the world's largest wireless network emulator with hardware-in-the-loop, to confirm the viability of our approach in an experimental network based on the open-source \gls{oai} 5G protocol stack~\cite{moro2023toward}.
\end{itemize}

We believe that the proposed approaches and test methodology is an important step toward the practical management and optimization of scalable, resilient, high-performance wireless self-backhaul solutions in realistic deployments.

\section{State of The Art}
\label{sec:soa}

Tezergil et al. recently surveyed the theme of planning wireless 5G backhaul networks \cite{tezergil2022wireless}. \rev{In this section, we review the papers with assumptions comparable to those used in our paper.}

Saha and Dhillon~\cite{saha2019milliter} use stochastic geometry to derive general results; however, their approach is intrinsically limited to 2-hop backhaul paths and relies on Poisson point process-based deployments, which do not include a real-world deployment assessment, unlike our proposed approach which includes real-world data validation. 
Lai et al. \cite{lai2020resource} present results based on random node placement, assuming a-priori knowledge of users' positions. In contrast, our approach does not require such assumptions, making it more broadly applicable. 
Madapatha et al. \cite{madapatha2021topology}, Pagin et al. \cite{pagin2022resource}, and Yuan et al. \cite{yuan2020optimal} focus on optimizing backhaul topology and scheduling once the \iabds are already placed. This presents a less generic and easier-to-solve problem compared to the joint optimization of placement and backhaul links that we tackle in our work.
Zhang et al. \cite{zhang2020cooptimizing} optimize scheduling fairness; Gopalam et al. \cite{gopalam22distributed} develop a distributed algorithm for scheduling optimization; Ma et al. \cite{ma2020qos} jointly optimize the scheduling of access and backhaul links; and Zhang et al. \cite{zhang2020joint} optimize spectrum and power allocation. All these works, however, operate on a fixed network topology, whereas our work jointly optimizes both topology and backhaul scheduling, adding flexibility to the solution.
Islam et al. \cite{islam2017integrated,islam2018investigation} propose an \gls{ilp} model for optimizing the placement of \iabds and link allocations to maximize flow, without minimizing cost while guaranteeing flow. In contrast, our approach also incorporates robustness, a critical factor for practical deployment, which is absent from their model. 
Mcmenamy et al. \cite{mcmenamy2020hop} employ a similar approach by limiting the number of hops towards terminals before allocating links among \gls{iab} nodes. However, their method is suboptimal, whereas we demonstrate that our approach achieves optimality in more than 90\% of real networks and near-optimal results in the remaining cases.

Like our work, these papers assume that backhaul links that do not share an edge are non-interfering due to the use of mmWave links and highly directive beamforming.

In summary, our work is original because it: (i) jointly addresses the problem of \gls{iab}-donor placement and backhaul optimization; (ii) tackles both topology and flow optimization together; (iii) introduces robustness into the backhaul; (iv) provides an optimal, not heuristic, solution; (v) uses real data and testbeds rather than random node placements in empty scenarios; and (vi) delivers a working prototype on a realistic testbed integrated into the O-RAN architecture.

\section{\gls{iab} Optimization --- Problem Statement}
\label{sec:problem-statement}

\begin{figure}
    \centering
    \includegraphics[width=\linewidth]{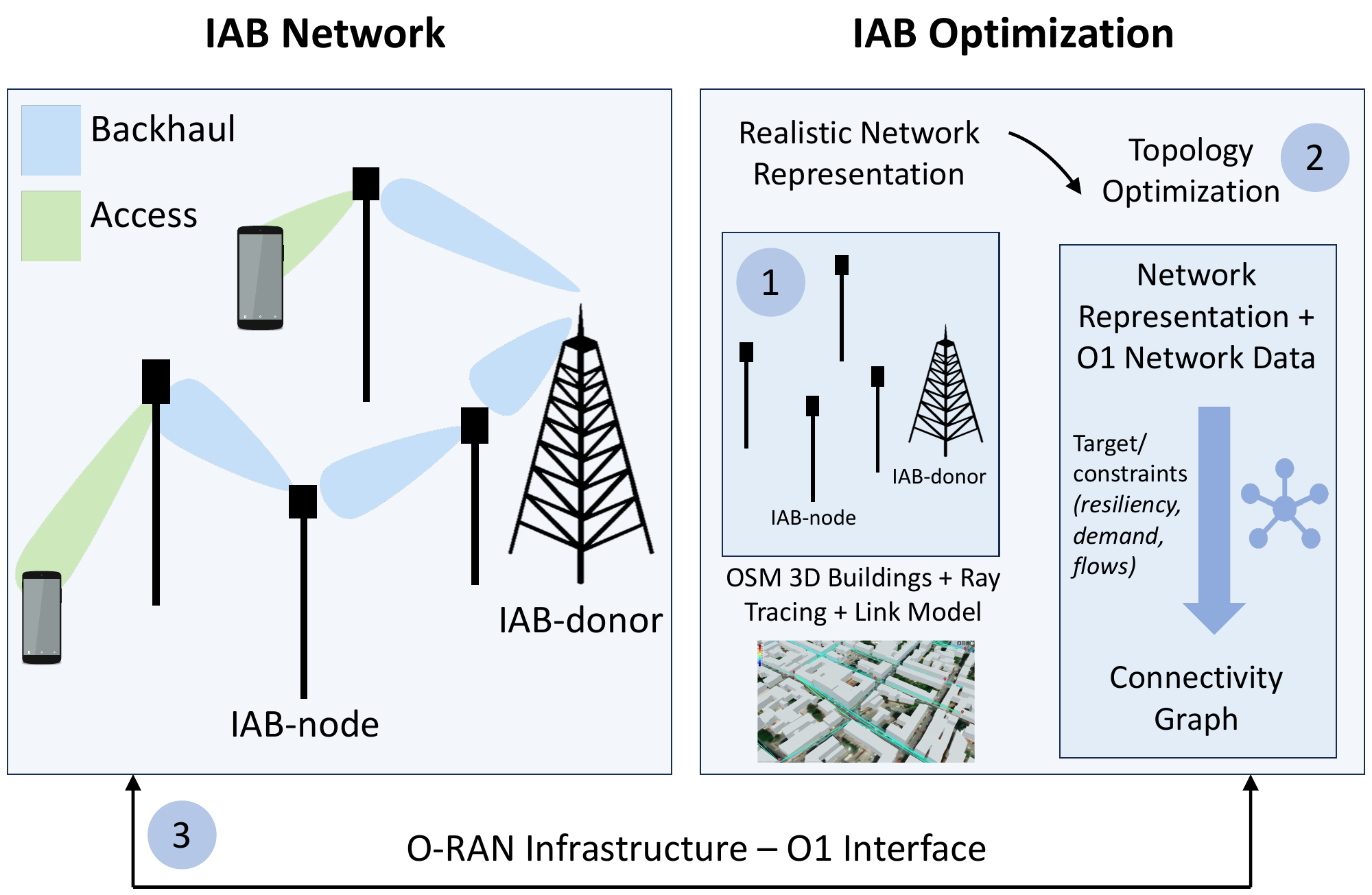}
    \caption{System model of the \gls{iab} closed-loop management optimization using the (i) realistic topologies representations; (ii) graph-based optimization; and (iii) the O-RAN infrastructure.}
    \label{fig:model}
\end{figure}

This section introduces the proposed system model for the design of optimal, resilient, and reconfigurable \gls{iab} topologies. Our solution pushes forward the state of the art and is practically implementable on available hardware and present standards. \Cref{fig:model} depicts the building blocks we rely on. First, we leverage realistic representations of the areas where the \gls{iab} network operates. Second, we develop optimization techniques that deploy optimal connectivity graphs over such topologies, including redundant paths to enforce network resiliency. Third, we leverage the O-RAN architecture, and, specifically, the O1 interface between the \ran and the \nonrt \ric, to perform reconfigurations of the network in case of failures.

\subsection{Realistic Deployment Area Representation}
\label{sec:topo}

We start from a 3D surface representing with high fidelity the deployment area.
This approach has been recently adopted \rev{in various papers}, e.g., to study localization \cite{villen2022prediction}, network planning \cite{gemmi2022cost}, and LoS estimation \cite{alhourani2020probability}. We use open data from public administrations and the OpenStreetMap (OSM) buildings project \cite{OSMbuildings}.
In this paper we use the methodology and heuristic proposed in \cite{gemmi2022cost} to deploy a network of \gnbs in an outdoor urban area, and we then evaluate LoS and link capacity between the \gnb similarly to \cite{alhourani2020probability}.
Given a certain urban scenario and a desired density of \gnb per square kilometer, we identify the positions of the \gnbs that provide optimal outdoor coverage and the availability of links between \gnbs.

We also leverage a realistic traffic demand profile, based on the knowledge of the area covered by each  \gnb and requirements for next-generation wireless systems, as we detail in \cref{sec:demand}, and a link model based on parameters for \gls{3gpp} systems and the \gls{oai} reference implementation, as discussed in \cref{sec:capacity}. In the remainder of the paper, we focus on a \gls{iab} network deployed at \glspl{mmwave}, to evaluate the impact of large bandwidth availability to the performance of the system.

\subsection{Topology Optimization}
\label{sec:optimization}

The optimization uses the given positions of \gnbs and creates an optimal \gls{iab} backhaul with the minimum number of \gls{iab}-donors, which require fiber connectivity to the core network. We follow the \gls{iab} specifications that support a backhaul network made of multiple loop-free trees rooted in \gls{iab}-donors. 
The optimization is performed before deploying the network, as it determines the nodes with wired access and those using a wireless backhaul. However, it produces redundant topologies that can be reconfigured at run-time to repair failures.

\begin{figure}
    \centering
    \includegraphics[width=.8\columnwidth]{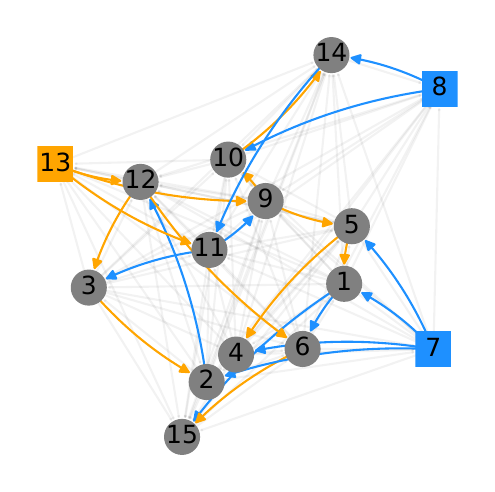}
    \caption{An example realization of the backhaul graph with $\red=2$. Circles are \iabns, squares are \iabds, orange/blue edges are backhaul links in two disjoint edge-set, gray edges are all the potential edges. Every gray node has two incoming edges.}\label{fig:multitree}
\end{figure}

We consider a directed network graph \vgraph, such as the one shown in \cref{fig:multitree}, that represents the backhaul network, made of a set  \vnset of \gnbs and a set \veset of potential edges, i.e., edges that can be used for the backhaul.
The first goal of the optimization is to find the smallest number of \iabds, i.e., the smallest subset $\bnset\subseteq\vnset$ that must be connected to the core network with a wired connection.
The second goal is to choose the subset $\beset\subseteq\veset$ of edges that create a multi-hop path from every \gnbs in $\vnset$ to some \iabd.  The resulting topology must respect some performance and reliability requirements. 
Altogether, the final goal is to define $\bnset$ and a new graph \bgraph made of all the \gnbs and the set \beset of edges resulting from the union of all the edges of all the trees. 
We will impose two classes of constraints: topological constraints that impose reliability features and flow conservation based on the estimated link capacity and traffic demand of every \gls{gnb}.
Reliability is one of the key features of 5G, and our approach can be parameterized in order to provide the desired level of redundancy.

\subsection{Open RAN for Optimized \gls{iab} Deployments}\label{sec:rapp_design}

This generated graph is used to optimally deploy the \gls{iab} network in the first instance, but also to manage the network through primitives and capabilities provided by the O-RAN architecture depicted in \cref{fig:model}. O-RAN introduced the \nonrt \ric, which performs optimization and service provisioning with a closed-loop control with a granularity higher than 1s, and connects to the \gls{ran} through the O1 interface. The control logic is defined by custom applications called rApps~\cite{oran-wg2-non-rt-ric-architecture}. The O-RAN architecture has been recently extended to support \gls{iab} operations, with the interfaces to the \glspl{ric} (e.g., O1) implemented as tunnels over the backhaul network~\cite{moro2023toward}. 

The \gls{iab} rApp is used in two phases. For network setup, it configures each network component based on the optimized topology. Then, it continuously receives performance metrics and failure events through the O1 interface, and uses them to detect relevant changes in the network topology (e.g., the degradation of a radio link). The rApp reacts by reconfiguring the \gls{iab} topology based on the optimization output. 
We prototype the rApp on Colosseum, a wireless network emulator that includes 256 \gls{sdr} together with a digital channel emulator to run large-scale experiments~\cite{bonati2021colosseum}, and showcase how network reconfigurability can be achieved. 

\section{Estimating Demand and Link Capacity}
\label{sec:link}

The starting point of our analysis is a 3D model of existing urban areas (obtained using open data) over which we apply state of the art algorithms to decide a placement of \gnbs that can guarantee the best ground coverage using LoS links \cite{gemmi2022cost}. We then need to estimate the traffic demand for every \gnb, and the link capacity that every point-to-point link between \glspl{gnb} can offer.  This will create the annotated graph \vgraph over which we run the optimization. Both estimations are data-driven and are part of our original methodology. 

\subsection{Estimating Demand}
\label{sec:demand}

The ITU provides precise parameters to simulate \gnb deployments when offering ``extended mobile broadband'' services \cite{itu2017guidelines}, which correspond to 10 \gls{ue} per \gnb, each one with a demand of at least 100 Mb/s, so each \gnb should be able to serve a load $\lambda = 1000$ Mb/s. This work focuses on the coverage of outdoor public areas, so assigning a fixed load per \gnb is realistic only if the goal is to minimize coverage overlap in a context where \glspl{gnb} can be deployed without constraints.
However, in scenarios based on actual topographies, as those used in this paper, the buildings make the area of interest nonhomogeneous, with some \glspl{gnb} that are placed in positions that cover specific areas that would otherwise be significantly shadowed~\cite{madapatha2023constrained}. As a result, coverage areas are often partially overlapping and the \glspl{ue} may be shared among \gnbs. This calls for a strategy to model shared demand among \glspl{gnb}.

We sample the area under analysis with one point per square meter. From the initial placement algorithm, we obtain $\Sigma = \{\sigma^0\ldots\sigma^n\}$ that is a family of sets. $\sigma^i$ includes all the discrete $(x,y)$ coordinates on the ground that are covered by \gnb $i\in\vnset$. We restrict $\sigma^i$ to the points on the ground that are in \los with the \gnb $i$, as we focus on a mmWave deployment, but the same approach can be used with a different definition (i.e., all the points with a minimum estimated capacity).  The area occupied by $\sigma^i$ is then simply $|\sigma^i|$ (the number of points in the sampled 2D space). One point can belong to more than one set, so we define the multiplicity $m_{x,y} = |\{\sigma^i \in \Sigma \;|\; (x,y)\in \sigma^i\}|$. We define the demand $\demi$ of a \gnb as:

\begin{equation}
\demi = \frac{|\sigma^i| \lambda}{\sum_{(x,y)\in \sigma^i} m_{x,y}}
\label{eq:demand}
\end{equation}
Note that if $\sigma^i$ is not overlapping with any other $\sigma^j$, then $m_{x,y} = 1 \; \forall \; (x,y)\in \sigma^i$ and $d_i = \lambda$, while if \gnb $i$ covers an area that overlaps with some other \gnb (as it happens for the majority of the real-world scenarios) then $d_i < \lambda$.

\subsection{Link Capacity Estimation}
\label{sec:capacity}
To estimate the capacity of each link between a pair of nodes $(s,d)$, we model the propagation through a ray tracing analysis using the MATLAB suite, and combine a link abstraction model based on the physical layer implementation of \gls{oai}. First, we load the 3D model of the buildings obtained from OSM Buildings, then for each pair of \gnbs we perform ray tracing using the shooting and bouncing method \cite{ling89shooting}. We consider up to a maximum of 4 reflections and we ignore the effect of diffraction that is negligible at mmWave frequencies~\cite{lecci2021accuracy}.

For each pair $(s,d)$ we obtain a set $\Gamma$ of rays, each ray $r$ is associated with a pathloss ($P_{r}$), a phase $\Phi_{r}$, a delay ($\delta_{r}$), the angle of arrival and the angle of departure of the ray ($AoA_{r}$, and $AoD_{r}$).
Each \gnb uses with a uniform planar array antenna, with $8\times8$ isotropic antenna elements spaced at half wavelength distance. 
The channel matrix $\mathbf{H}$ is~\cite{lecci2021accuracy}
\begin{equation}
    \mathbf{H} = \sum_{r=1}^\Gamma \sqrt{P_{r}}~ e^{j(-2\pi\delta_{r}f_c+\Phi_{r}})~ \mathbf{a}_{rx}^\star (AoA_{r}) ~\mathbf{a}_{tx}^H(AoD_{r}),
\end{equation}
where $\mathbf{a}_{rx}$ and $\mathbf{a}_{tx}$ are respectively the receiver and transmitter array responses, $^\star$ is the conjugate operator and $^H$ is the hermitian operator. We compute $\mathbf{H}$ for every pair of \gnbs $(s,d)$, and derive the \gls{snr} $S_{s,d}$ as:
\begin{equation}
    S_{s,d} = \frac{\Pi \mathbf{w}_{s,d}^T \mathbf{H}_{s,d} \mathbf{w}_{d,s}}{N_0BN_f},
\end{equation}
where $\Pi$ is the transmission power and $\mathbf{w}_{s,d}$ ($\mathbf{w}_{d,s}$) is the beamforming vector used by the device $s$ ($d$) to communicate with the device $d$ ($s$), obtained by applying singular value decomposition on the channel matrix $\mathbf{H}_{s,d}$. At the denominator, $N_0$ is the noise density in W/Hz, $B$ is the bandwidth, and $N_f$ is the noise figure. 
The values of the parameters are reported in \cref{table:simulation_params}. 
As already mentioned in \cref{sec:soa}, we assume that backhaul links do not interfere with each other, thanks to large MIMO arrays and highly directive links, and use the \gls{snr} to model the link quality with a 5G link abstraction model.

From the open-source \gls{oai} 5G \gls{ran} implementation~\cite{kaltenberger2020openairinterface}, we extract the table of triplets $(S_i, M_i, E_i)$ that matches a certain SNR with a given \gls{mcs} and a \gls{bler} \cite{OAITable}.
Then we can compute the maximum MCS $M_{s,r}$ as follows:
\begin{equation}
M_{s,r} = \max {M_i} ~~ \text{s.t.} ~~ E_i < 0.1 ~~ \text{and} ~~ S_{s,r} \geq S_i.
\end{equation}
The downlink capacity $C$ is then
\begin{equation}
C_{s,d} =   \Lambda Q(M_{s,r}) R(M_{s,r}) \frac{12 RB}{T_u} (1-Oh) R_{slot},
\end{equation}
where $\Lambda$ is the number of MIMO layers, $Q(M)$ and $R(M)$ are two functions associating the MCS to the modulation order and the code rate, $RB$ is the number of Resource Blocks used, $Oh$ is the control channel overhead, $R_{slot}$  is the ratio of Downlink to Uplink slots used, and $T_u$ is the average duration of an OFDM symbol. 
Further details regarding the formula and the different values can be found in the 3GPP technical specifications \cite{3gpp.38.306}.

\begin{table}
    \centering
    \begin{tabular}{lll}
        \toprule
        Parameter & Value \\
        \midrule
        Carrier frequency ($f_c$) & \SI{27}{\giga\hertz} \\
        Bandwidth ($B$) & \SI{400}{\mega\hertz} \\
        Resource Blocks ($RB$) & 132 \\
        Numerology ($\mu$) & 3 \\
        Uplink to downlink slot ratio ($R_{slot}$) & 0.7 \\
        Control channel overhead ($Oh$) & 0.18 \\
		Noise density ($N_0$)  & \SI{-174}{\deci\bel\m/\hertz}\\
        Noise Figure  ($N_f$) & \SI{7}{\deci\bel} \\
        Antenna Elements & 8x8 \\
        Maximum number of reflection & 4 \\
        MIMO Layers ($\Lambda$) & [1,2]\\
        Transmission power ($\Pi$) & \SI{33}{\deci\bel\m} \\ 
        Maximum BLER & 0.1\\ 
        \bottomrule
    \end{tabular}
    \caption{Simulation parameters. Most parameters are adapted from~\cite{itu2017guidelines}.}
    \label{table:simulation_params}
\end{table}

\section{Optimization Models}
\label{sec:model}

\rev{The methodology described in the previous two sections, enable us to optimize the topology of the \gls{iab} network with different objectives and reliability constraints. In this section we will provide the details on the different optimization models.}

Let us consider the edges in \beset as directed, representing the downstream links from the donors down to the \gnbs. We focus on downstream for simplicity but the problem can be extended to both downstream and upstream. 
Given an \iabn $i$ in a certain tree, let the out-degree $i$ be its number of direct children and the distance of $i$ be the number of hops to the \gls{iab}-donor.
Next, we first define the topological constraints to produce a robust graph, and then add flow constraints. Parameters that are common to both models are (i)\D, the maximum distance from a \gnb to an \gls{iab}-donor, which can be set to limit the delay; (ii) \dg, the out-degree of a \gnb; (iii) \eij, a set of parameters so that $\eij=1$ if and only if the edge from $i$ to $j$ is present in \veset, i.e., the SNR is sufficient to negotiate a link.

\begin{figure}
    \centering
    \includegraphics[width=.8\columnwidth]{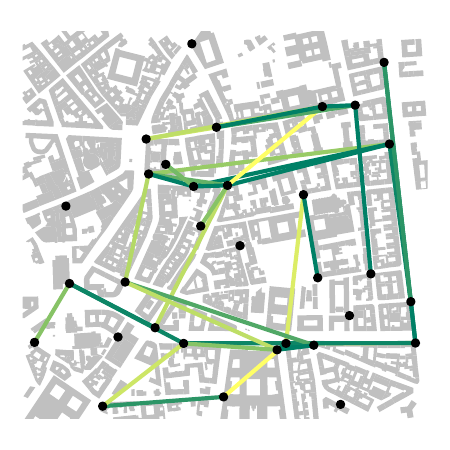}
    \caption{Map showing one of the four realistic networks in Milan, with density 45 \glspl{gnb}/km$^2$. \gls{iab}-nodes are in black, and feasible backhaul links have a color gradient corresponding to their capacity (yellow for low capacity and green for high capacity).}\label{fig:realistic_tree}
\end{figure}

\subsection{Topological Constraints}
We divide the set of edges in \red disjoint sets, and impose that every tree rooted in an \gls{iab}-donor must use edges coming from only one out of $\red$ sets, while every \gls{iab}-node must belong to at least \red trees. If $\red=2$, there are two separate trees that serve each \gls{iab}-node $i$: if one edge fails in one tree and $i$ remains isolated, the second tree can be used to serve \gls{iab}-node $i$. This results in $\red$ disjoint sets of edges, however we may have more trees, because the topological constraints and the availability of edges may make it impossible to have only $R$ trees. Each tree will have disjoint edges from each other tree, and each \gls{iab}-node will have \red parents, belonging to \red different trees rooted in two different \gls{iab}-donors.
\Cref{fig:multitree} shows an example of an optimized topology when $D=3$ and $\red=2$. The optimization creates two separate edge-set, represented using two colors. The optimization yields 3 \gls{iab}-donors out of 18 total \glspl{gnb} (thus 15 \gls{iab}-nodes), and each \gls{iab}-donor is the root of a tree using edges coming only from one set. Every \gls{iab}-node has two parents connected with links using two distinct colors. As a consequence, if one edge fails and an \gls{iab}-node can not reach the \gls{iab}-donor at the root of the tree using the blue edge-set, it can still reach a \gls{iab}-donor using the tree made of orange edges. 

The binary variable \uilk is set to 1 if \gls{iab}-node $i$ is at distance $l$ from the donor that is the root of a tree with edges in the sub-set $k\in [1,\ldots\red]$. The binary variable \pijk is set to 1 if the edge from $i$ to $j$ is active in a tree that belongs to sub-set $k$.
The optimization minimizes the total number of donors:
\begin{equation}
\text{objective:} \quad \text{min}\sum_{i\in \vnset}\sum_{k=1}^{R} u_{i,0,k} \label{eq:obj},
\end{equation}
under the following constraints:
\begin{align}
\sum_{l=0}^\D \uilk = 1 - \sum_{r\neq k}u_{i,0,r} \quad \forall i, \forall k \label{c:rdist}\\
\sum_k u_{i,0,k} \leq 1 \quad \forall i \label{c:rsingleroot}\\
\sum_{i\in \vnset} \pijk = 1 - \sum_r u_{j,0,r} \quad \forall j,k \label{c:rdonor}\\
\sum_{j\in \vnset} \pijk < \dg \quad \forall i,k \label{c:rdeg}\\
\pijk \leq 1 - u_{j,l,k} + u_{i, l-1,k} \quad \forall i,j,k \quad \forall l\in\{1\ldots\D\} \label{c:rpath}\\
\pijk \leq \eij \quad \forall i,j,k \label{c:rexist}\\
\sum_k \pijk + \sum_k P_{j,i,k} \leq 1 \quad \forall i,j. \label{c:rdir}
\end{align}
\cref{c:rdist} imposes that every \gnb must belong to at least one tree per edge-set $k$, unless it is an \gls{iab}-donor, which is the root of exactly one tree; \cref{c:rsingleroot} imposes that each \gls{iab}-donor must be the root of a single tree; 
 \cref{c:rdonor} imposes that one \gls{iab}-donor of a certain tree has no incoming edges, and other nodes have one (thus it is a tree topology); 
\cref{c:rdeg} imposes the maximum out-degree; 
\cref{c:rpath} imposes consistency on paths, hop by hop, that is, in a certain tree $k$ if a link connecting \gls{iab}-node $i$ to \gls{iab}-node $j$ is active, then the distance of \gls{iab}-node $i$ equals the distance of \gls{iab}-node $j$ minus one;
 \cref{c:rexist} imposes that only existing links can be used; \cref{c:rdir} imposes that every edge is used only in one direction and only in one tree. This last equation produces trees that are edge-disjoint.

There is always one trivial acceptable solution in which all \gnbs are \gls{iab}-donors and the number of variables scales as $\red|\vnset|^2$ as it is dominated by the dimension of \pijk.

\subsection{Flow Constraints}

We now introduce the flow-based constraints, implemented as a set of new equations on top of the topology-based ones, thus producing a topology and flow-based model. The flow-based model introduces three new sets of parameters, which are \demi (the flow demand for \gls{iab}-node $i$), \capi (the total maximum outgoing flow for an \gls{iab}-donor), and \capij (the link capacity of \eij). 
It also introduces two new sets of variables. \fijhk represents the flow destined to \gls{iab}-node $h$ passing through edge $e_{i,j}$ belonging to edge-set $k$. This is required because even if edges can not belong to more than one tree, the flow constraints need to be guaranteed in each edge and in every tree.
\luij is a real value in $[0,1]$ that represents the fraction of time link $e_{i,j}$ is used. As each node $j$ has a single radio for access and backhaul, it becomes necessary to impose a constraint on the total number of active links for $j$, modeled through variable $a_{i,j}$. If $j$ is connected to, for example, node 1 in upstream and node 2 in downstream, then $a_{1,j} + a_{2,j} \leq 1$.

The flow constraints are set as follows:
\begin{align}
f_{i,j,i,k} = 0 \quad \forall \; i,j,k \label{eq:noselfb}\\
\sum_{h}\sum_j\fijhk - \sum_{h\neq i}\sum_j f_{j,i,h,k} \leq \capi \sum_k u_{i,0,k} \quad \forall i,k \label{eq:flowconb}\\
\sum_j f_{j,i,i,k} \geq (1-\sum_h u_{i,0,h})\demi\label{eq:incflowb} \quad \forall i,k\\
\luij \leq \sum_k\pijk \label{eq:maxusageb} \quad \forall \; i,j\\
\sum_i\sum_h\sum_k \fijhk \leq \sum_i \capij\luij \label{eq:maxflowpernodeb} \forall j\\
\sum_h\sum_k \fijhk \leq \luij \capij  \quad \forall i,j \label{eq:maxflowlinkb}\\
\fijhk \leq \capij \pijk \quad \forall i,j,h,k  \label{eq:maxlinkflowb}
\end{align}

Here, \cref{eq:noselfb} prevents self-loops in the flow graph, \cref{eq:flowconb} imposes flow conservation for all \gls{iab}-nodes and all edge-set, except for donors that can output a flow \capi. Note that the real flow outgoing node $i$ is capped in every link by \cref{eq:maxflowlinkb}, so \capi is just an arbitrary non zero upper bound that makes the equation formally correct, and it can model the bit-rate available on the wired link. \cref{eq:incflowb} imposes that every \gls{iab}-node receives at least the amount of flow corresponding to its demand; 
\cref{eq:maxusageb} imposes that the usage on a link must be zero if the link is not chosen by the topology optimization;  \cref{eq:maxflowpernodeb} imposes that an \gls{iab}-node can not receive in input more flow that what all its incoming link allow; similarly \cref{eq:maxflowlinkb} imposes that flow on a link does not exceed its allocated capacity. 
\cref{eq:maxlinkflowb} imposes that if an edge is not assigned to tree $k$ the corresponding flow in tree $k$ is zero.
The objective function is \cref{eq:obj}, as these conditions only impose more constraints on the \gls{iab}-donors, with still a trivial solution with all \gnbs as \gls{iab}-donors. 

The cost associated with the model is the added complexity: the number of variables scales as $|\vnset|^3$ as it is dominated by the dimension of \fijh. However, since every \gls{iab}-node has only one parent if the capacity of one incoming link to \gls{iab}-node $j$ is lower than the demand of \gls{iab}-node $j$ ($L_{i,j} < d_j$), we can set $e_{i,j} = 0$ as that link cannot serve the demand of $d_j$. This allows us to prune some edges in $\veset$ before running the optimization.

Finally, while our model is designed for a greenfield deployment, in which the operator is planning the network from scratch, it can also be used in a brownfield deployment in which some \gls{iab}-donors are already connected to the core with a fiber connection. The only required change in the model is the need to force some of the $u_{i,0,k}$ variables to be constant set to 1. 
This is important because we can use the optimization for both new networks, or to upgrade or dynamically control existing ones, as shown in \cref{sec:rapp}. With a similar reasoning, if some of the locations of the \gnbs is preferable than others, for instance if the cost of connecting it with fiber is smaller, the objective function can be modified so that every \gnb has a different cost and not a simple unitary cost as in \cref{eq:obj}.

The whole problem is based on a multi-commodity flow problem merged with a shortest path multi-tree problem. There is however a key difference, the number and position of sources of the commodity are not decided a priori, but must be decided by the optimization. To the best of our knowledge, this specific problem has not been so far formulated. The problem is clearly non-polynomial because it mixes the multi-commodity flow with a combinatorial minimization of all the possible set of sources of the commodity. We also give the network designer the freedom to choose a maximum out the degree of and the distance from the \gls{iab}-donors. The second parameter in particular will affect the network delay, so it is important that the operator can define it based on the specific applications that must be supported.

\section{Numerical Results}
\label{sec:num-results}

In this section, we report results on the feasibility and the effectiveness of the proposed optimization, leveraging the resilient and non-resilient versions of the flow optimization problem. We use as a metric the fraction $\rho$ of \gls{iab}-donors in the network, which is the ratio between the value of the objective function and the total size of the network $|\vnset|$:
\[
\rho = \frac{\sum_{i\in \vnset}\sum_{k=1}^{R} u_{i,0,k}}{|\vnset|}.
\]
First, we consider synthetic random graphs, to verify the feasibility of our approach in a controlled scenario, and then realistic graphs generated from open data of four European cities.

\subsection{Synthetic Graphs}

The synthetic graphs are generated by placing nodes in random positions in a 2D area, with an average density of 45 \glspl{gnb}/km$^2$, which achieves 95\% coverage of the outdoor urban areas~\cite{gemmi2022cost}. We vary the area to deploy 15, 30, and 45 \glspl{gnb}. The demand is estimated with \cref{eq:demand}, assuming a coverage radius of 100 m. We use the 3GPP TR 38.901 technical report for modeling both the probability of \los between two \glspl{gnb} and the path loss~\cite{3gpp.38.901}. We thus do not explicitly model obstacles, and obtain networks with an edge density substantially higher than in real settings. We compute the link capacity as described in \cref{sec:capacity}, using the parameters in \cref{table:simulation_params} with 1 and 2 MIMO layers $\Lambda$. Execution times are evaluated on a 16 cores server (Intel Xeon Gold 6342 CPU @ 2.80GHz), with 64GB of RAM using the Gurobi solver with 30 randomly generated graphs for each number of \glspl{gnb}. After 48 hours the solver was stopped, and for those runs that did not reach the optimal value, we report the upper bound distance from the optimum. 

\begin{figure}
         \centering
         \definecolor{c1}{RGB}{213,94,0}
\definecolor{c2}{RGB}{1,115,178}
\definecolor{c3}{RGB}{2,158,115}
\definecolor{c4}{RGB}{176,176,176}
\definecolor{c5}{RGB}{222,143,5}
\definecolor{c6}{RGB}{204,204,204}
\definecolor{c7}{RGB}{204,120,188}
\pgfplotstableset{col sep=comma}
\pgfplotsset{
    only if/.style args={entry of #1 is #2}{
        /pgfplots/boxplot/data filter/.append code={
            \edef\tempa{\thisrow{#1}}
            \edef\tempb{#2}
            \ifx\tempa\tempb
            \else
            \def\pgfmathresult{}
            \fi
        }
    }
}

\pgfplotsset{
    compat=1.15,
    my boxplot style/.style={
        width=.51\linewidth,
        height=.5\linewidth,
        ylabel near ticks,
        xticklabel style = {align=center, font=\scriptsize},
        ylabel style ={font=\footnotesize},
        xlabel style ={font=\footnotesize},
        y grid style={c4, draw opacity=0.5},
        xmajorgrids,
        ymajorgrids,
        ytick style={color=black},
        boxplot,
        boxplot/draw direction=y,
        ymin=0.1, ymax=0.9,
        major tick length=2pt
    },
}

\subfloat[$\Lambda=1,\, R=1$]{
    \begin{tikzpicture}
        \begin{axis}[
            my boxplot style,
            xtick = {1,2,3},
            xticklabels = {15,30,45},
            xlabel={$|\vnset|$},
            ylabel={$\rho$},
            yticklabel style = {align=center, font=\scriptsize},
        ]
        \foreach \n in {15,30,45} {
            \addplot [draw=black] table 
                [y = donor_ratio,
                only if={entry of l2model is single},
                only if={entry of MIMOL is 1.0},
                only if={entry of n is \n},
                ] {data/synth-results.csv};
            }
        \end{axis}

    \end{tikzpicture}

}
\subfloat[$\Lambda=2,\, R=1$]{
    \begin{tikzpicture}
        \begin{axis}[
            my boxplot style,
            xtick = {1,2,3},
            xticklabels = {15,30,45},
            xlabel={$|\vnset|$},
            yticklabel=\empty,
        ]
        \foreach \n in {15,30,45} {
            \addplot [draw=black] table 
                [y = donor_ratio,
                only if={entry of l2model is single},
                only if={entry of MIMOL is 2.0},
                only if={entry of n is \n},
                ] {data/synth-results.csv};
            }
        \end{axis}
    \end{tikzpicture}
}\\

\subfloat[$\Lambda=1, R=2$]{
    \begin{tikzpicture}
        \begin{axis}[
            my boxplot style,
            xtick = {1,2,3},
            xticklabels = {15,30,45},
            xlabel={$|\vnset|$},
            ylabel={$\rho$},
            yticklabel style = {align=center, font=\scriptsize},
        ]
        \foreach \n in {15,30,45} {
            \addplot [draw=black] table 
                [y = donor_ratio,
                only if={entry of l2model is multitree},
                only if={entry of MIMOL is 1.0},
                only if={entry of n is \n},
                ] {data/synth-results.csv};
            }
        \end{axis}        
    \end{tikzpicture}

}
\subfloat[$\Lambda=2, R=2$]{
    \begin{tikzpicture}      
        \begin{axis}[
            my boxplot style,
            xtick = {1,2,3},
            xticklabels = {15,30,45},
            xlabel={$|\vnset|$},
            yticklabel=\empty,
        ]
        \foreach \n in {15,30,45} {
            \addplot [draw=black] table 
                [y = donor_ratio,
                only if={entry of l2model is multitree},
                only if={entry of MIMOL is 2.0},
                only if={entry of n is \n},
                ] {data/synth-results.csv};
            }
        \end{axis}
    \end{tikzpicture}
}
         \caption{Box-plots of the fraction of donors $\rho$ for the synthetic topologies, with different MIMO configuration ($\Lambda=1$, $\Lambda=2$), the single tree model ($R=1$) and the failure resistant model ($R=2$). The box plot show the median, 25\% and 75\% quartile and 1.5*IQR (inter quantile range) whiskers.}
         \label{fig:poisson}
\end{figure}
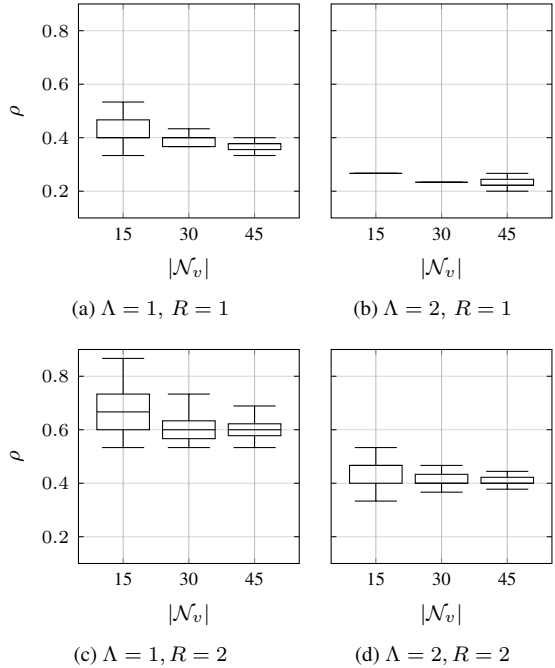

\Cref{fig:poisson} shows the ratio $\rho$ between \gls{iab}-donors and total number of \glspl{gnb} in the simulation. The two leftmost plots (a, b) report the results with only one tree ($R=1$), while plots (c, d) report the failure resilient model ($R=2$), both with $\Lambda=1$, $\Lambda=2$.
As a general trend, higher $\Lambda$ and lower $R$ lead to lower $\rho$. This is expected since with a higher capacity per link the backhaul network can transport more traffic to a smaller number of donors, while redundancy requires more independent trees and thus more donors. Increasing $|\vnset|$ enlarges the space of possible results and can lead to a further reduction of the number of required \gls{iab}-donors. For the most challenging configuration ($\Lambda=1, R=2$), the median value of $\rho$ ranges from 0.66 to 0.6, which means that we can save up to 40\% of the donors. Savings increase to up to 60\% with $\Lambda=2, R=2$ and 45 \gls{gnb}/km$^2$, which means that 27 \glspl{gnb} out or 45 do not need to be connected with a fiber cable.

The failure resilient topology imposes the strong condition that all flow is conserved and there is no performance degradation upon the failure of a link. If we relax this condition, and set $R=1$, we can further reduce the number of \iabds down to 37\% (45 \gnbs, $\Lambda=1$) and 23\% (45 \gnbs, $\Lambda=2$), introducing a non-zero probability of network outage. This does not necessarily imply that the failure of one link disconnects some \gls{ue}, but it could produce a certain performance degradation. Models can be tailored to provide a trade-off between these two approaches, obtaining $\rho$ close to the ones generated with $R=1$ with a predictable performance penalty in case of failure.

For solver execution runs that lasted at most 48 hours, a guaranteed optimal solution has been identified in 67\% of the cases. In the remaining ones, the upper bound of the distance of $\rho$ from the global optimum is on average 6.3\%. This is perfectly compatible with the network planning use case.

\subsection{Real-world Graphs}

The realistic graphs were generated for 4 cities (Florence and Milan in Italy, Barcelona in Spain, and Luxembourg city), in an area of about one km$^2$, as in \cref{fig:realistic_tree}.\footnote{As areas are selected based on street boundaries, they will not measure exactly 1 km$^2$, leading to slight variation in the number of \glspl{gnb} for each city.}

For each scenario, we use densities of 30, 45, and 60 \glspl{gnb}/km$^2$; run the placement heuristic to position the \glspl{gnb}; and compute the capacity as described in \cref{sec:capacity}. Some \glspl{gnb} are placed out of range from all others, and thus removed from the graph.

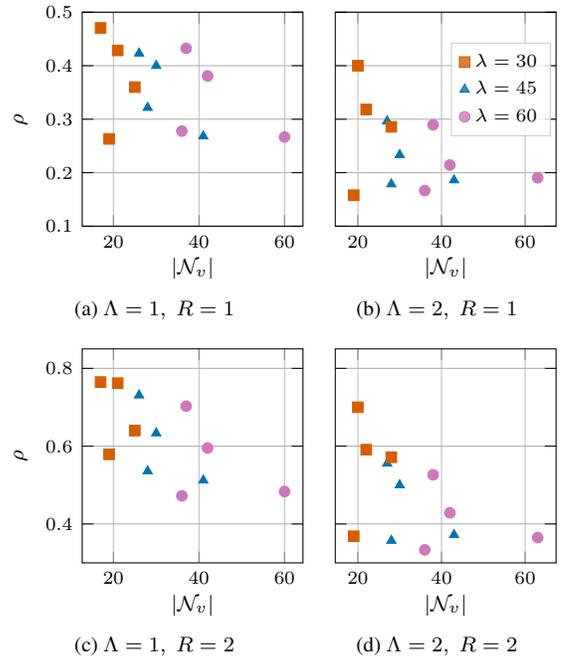
\begin{figure}
         \centering
        \definecolor{c1}{RGB}{213,94,0}
\definecolor{c2}{RGB}{1,115,178}
\definecolor{c3}{RGB}{2,158,115}
\definecolor{c4}{RGB}{176,176,176}
\definecolor{c5}{RGB}{222,143,5}
\definecolor{c6}{RGB}{204,204,204}
\definecolor{c7}{RGB}{204,120,188}
\pgfplotstableset{col sep=comma}
\pgfplotsset{
    discard if not/.style 2 args={
        x filter/.append code={
            \edef\tempa{\thisrow{#1}}
            \edef\tempb{#2}
            \ifx\tempa\tempb
            \else
                \def\pgfmathresult{inf}
            \fi
        }
    }
}
\pgfplotsset{
    compat=1.15,
    my scatter style/.style={
        width=.51\linewidth,
        height=.5\linewidth,
        ylabel near ticks,
        xticklabel style = {align=center, font=\scriptsize},
        yticklabel style = {align=center, font=\scriptsize},
        ylabel style ={font=\footnotesize},
        xlabel style ={font=\footnotesize},
        y grid style={c4},
        xmajorgrids,
        ymajorgrids,
        xlabel shift=-3pt,
        ytick style={color=black},
    },
}

\subfloat[$\Lambda=1,\; R=1$]{
\begin{tikzpicture}
  \begin{axis}[
    my scatter style,
    ymin=0.1, ymax=0.5,
    ylabel={$\rho$},
    xlabel={$|\vnset|$},
    ]
    \addplot [
        scatter,
        only marks,
        discard if not={l2model}{single},
        discard if not={MIMOL}{1x1},
        point meta=explicit symbolic,
        scatter/classes={
        30={mark=square*,c1},
        45={mark=triangle*,c2},
        60={mark=*,c7}
        },
        ] table [x=nodes, y = donor_ratio, meta=density] {data/real-results.csv};
    \end{axis}

\end{tikzpicture}
}
\subfloat[$\Lambda=2,\; R=1$]{
    \begin{tikzpicture}        
      \begin{axis}[
        my scatter style,
        ymin=0.1, ymax=0.5,
        xlabel={$|\vnset|$},
        yticklabel=\empty,
        legend cell align={left},
        legend columns=1,
        legend style={draw opacity=1, text opacity=1, draw=c6, fill opacity=0.8, at={(0.3,0.5)},anchor=west, font=\scriptsize, at={(0.75,0.42)}, anchor=south},
        ]
        \addplot [
            scatter,
            only marks,
            discard if not={l2model}{single},
            discard if not={MIMOL}{2x2},
            point meta=explicit symbolic,
            scatter/classes={
            30={mark=square*,c1},
            45={mark=triangle*,c2},
            60={mark=*,c7}
            },
            ] table [x=nodes, y = donor_ratio, meta=density] {data/real-results.csv};
            \legend{$\lambda=30$,$\lambda=45$,$\lambda=60$}
        \end{axis}

    \end{tikzpicture}

}\\
\subfloat[$\Lambda=1,\; R=2$]{
    \begin{tikzpicture}
      \begin{axis}[
        my scatter style,
        ylabel={$\rho$},
        xlabel={$|\vnset|$},
        ymin=0.3, ymax=0.85,
        ]
        \addplot [
            scatter,
            only marks,
            discard if not={l2model}{multitree},
            discard if not={MIMOL}{1x1},
            point meta=explicit symbolic,
            scatter/classes={
            30={mark=square*,c1},
            45={mark=triangle*,c2},
            60={mark=*,c7}
            },
            ] table [x=nodes, y = donor_ratio, meta=density] {data/real-results.csv};
        \end{axis}

    \end{tikzpicture}

}
\subfloat[$\Lambda=2,\; R=2$]{
    \begin{tikzpicture}
      \begin{axis}[
        my scatter style,
        xlabel={$|\vnset|$},
        ymin=0.3, ymax=0.85,
        yticklabel=\empty,
        ]
        \addplot [
            scatter,
            only marks,
            discard if not={l2model}{multitree},
            discard if not={MIMOL}{2x2},
            point meta=explicit symbolic,
            scatter/classes={
            30={mark=square*,c1},
            45={mark=triangle*,c2},
            60={mark=*,c7}
            },
            ] table [x=nodes, y = donor_ratio, meta=density] {data/real-results.csv};
        \end{axis}
    \end{tikzpicture}
}
         \caption{Fraction of donors $\rho$ for the realistic topologies, with different MIMO configuration ($\Lambda=1$, $\Lambda=2$), the single tree model ($R=1$) and the failure resilient model ($R=2$).}         \label{fig:scatter}
\end{figure}

\Cref{fig:scatter} confirms results obtained in the synthetic scenarios, i.e., also in the realistic scenario fewer \gls{iab}-donors are required as the density increases. In the realistic topologies the presence of obstacles makes the the topology less dense, and the distribution of users more concentrated, eventually this reduces the complexity of the problem and we can complete the optimization even with 60 \glspl{gnb}/km$^2$, and in some configurations $\rho$ is lower than 20\% (\cref{fig:scatter}(b)).
Only in 4 cases over 48 runs the optimization did not reach the optimal value, and in those 4 cases the upper bound of the distance of $\rho$ from the global optimum is on average 6\%.

Overall our results confirm that in both synthetic and real-world scenarios our optimization scheme is reliable and produces resilient topologies that can save a large number of \gls{iab}-donors, reducing the capital expenditure of fiber backhaul.

\section{Backhaul rApp Prototype and Validation}
\label{sec:rapp}

In this section, we discuss the integration of our methodology within the O-\gls{ran} framework, allowing operations and management of real-world networks. 
We implement a prototype rApp running in the \nonrt \gls{ric}, which processes a network made of $R=2$ trees (a primary and a back-up one) and gracefully handles the failure of a link by reconfiguring all the \gnbs to switch from their primary tree to the backup one. We validate the rApp using OpenAirInterface on Colosseum.

The application is initialized with the multitree topology obtained in the network planning phase. Once the network is set-up and running, the rApp maintains an updated representation of the \gls{iab} network using specific messages (HeartBeats, Performance Reports, and Fault Events) exchanged with the RAN through the O-RAN O1 Interface (see \cref{fig:model}). 
Upon the failure of a radio link between \gls{iab}-nodes, the upstream node, still connected to the core, detects the \gls{rrc} failure and sends a Failure Event message to the \gls{ric} through the O1 interface. The rApp reconfigures the network by updating the upstream connectivity of the \gls{iab}-node (which may require a reset of the stack in the OAI prototype). It also updates all the \gls{iab}-nodes downstream since they need to reconnect to the network core using a new path. Finally, it reconfigures routing within the core network to reflect the new downstream \iabns topology. 

\subsection{Validation on Colosseum}

\begin{figure}
    \centering
    \includegraphics[width=.7\linewidth]{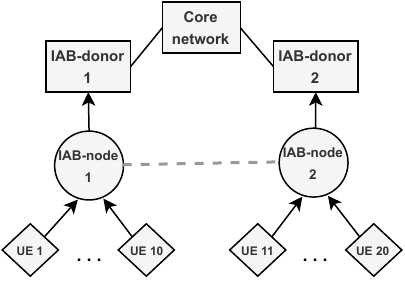}
    \caption{\gls{iab} topology including the Core network, two \gls{iab}-donors, two \gls{iab}-nodes and 20 \glspl{ue}. Black arrows indicate the normal topology and the grey dotted line indicates the backup link between the two \gls{iab}-nodes.}\label{fig:fail_topology}
\end{figure}

Currently, the \gls{oai} stack does not support handover between different upstream nodes, so it introduces long delays and performance fluctuations due to the need to restart the protocol stack. For this reason, our goal is not to provide an accurate performance analysis, but to demonstrate that our approach can be implemented in a real network. We have set up a representative \gls{iab}-network topology, shown in \cref{fig:fail_topology}, comprising two \gls{iab}-donors, two \gls{iab}-nodes and twenty \glspl{ue}. Each \iabn is connected to a different \iabd, and a feasible (but unused) link is available between the two \iabns. Note that we assume \gnbs have a secondary communication channel for their management through the O1 interface, possibly using the sub-GHz bands in order to be less subject to blockage~\cite{kwon2022resource}.

In our experiment, once the rApp initializes the whole network,
all \glspl{ue} and all \iabns start exchanging \gls{icmp} traffic with a server in the core network, routed through the parent \gls{iab}-node or donor.
Then, we emulate the failure of the link between \gls{iab}-node 2 and its donor. 
This triggers the transmission of a Fault Event message to the rApp, which in turn triggers the reconfiguration of \iabn 2, which will connect to \iabn 1 to reach the core. Since the current implementation of IAB for Colosseum and \gls{oai} creates end-to-end tunnels from \glspl{ue} to the core network~\cite{moro2023toward}, the \glspl{ue} attached to \iabn 2 must also be reconfigured. Moreover, the rApp reconfigures the 5G core, so that the backward path to the \glspl{ue} is also restored. 

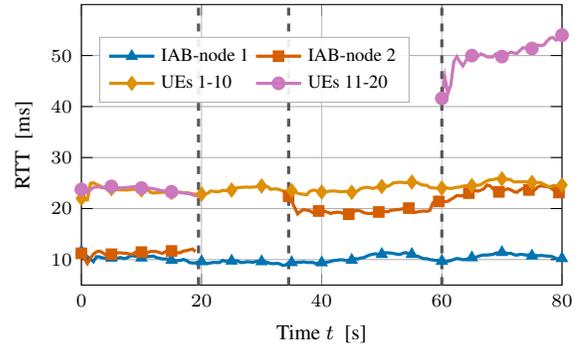
\begin{figure}
    \centering
    \definecolor{c1}{RGB}{213,94,0}
\definecolor{c2}{RGB}{1,115,178}
\definecolor{c3}{RGB}{2,158,115}
\definecolor{c4}{RGB}{176,176,176}
\definecolor{c5}{RGB}{222,143,5}
\definecolor{c6}{RGB}{204,204,204}
\definecolor{c7}{RGB}{204,120,188}
\pgfplotstableset{col sep=comma}

\pgfplotsset{
    compat=1.15,
    my boxplot style/.style={
        width=.9\linewidth,
        height=.6\linewidth,
        ylabel near ticks,
        xticklabel style = {align=center, font=\scriptsize},
        yticklabel style = {align=center, font=\scriptsize},
        ylabel style ={font=\footnotesize},
        xlabel style ={font=\footnotesize},
        y grid style={c4},
        xmajorgrids,
        ymajorgrids,
        ytick style={color=black},
        legend cell align={left},
        legend columns = 2,
        legend style={draw opacity=1, text opacity=1, draw=c4, fill opacity=0.9, font=\scriptsize, at={(0.36,0.65)}, anchor=south},
    },
}

\begin{tikzpicture}
    \begin{axis}[
        my boxplot style,
        xlabel={Time $t$ ~[s]},
        ylabel={RTT ~[ms]},
        unbounded coords=jump,
        ytick = {10,20,30,40,50},
        xmin = 0, xmax=80,
        ymin=5, ymax=60,
    ]
        \addplot [c2, very thick, mark=triangle*, mark repeat=10,mark options={solid,scale=0.8}] table 
        [   x = delta_time,
            y = iab_2,
            unbounded coords=jump,
        ] {data/failure.csv};
        \addlegendentry{\gls{iab}-node 1}
        \addplot [c1, very thick, mark=square*, mark repeat=10,mark options={solid,scale=0.8}] table 
        [   x = delta_time,
            y = iab_1,
            unbounded coords=jump,
        ] {data/failure.csv};
        \addlegendentry{\gls{iab}-node 2}
        \addplot [c5, very thick,mark=diamond*, mark repeat=10,mark options={solid,scale=1}] table 
        [   x = delta_time,
            y = ues_2,
        ] {data/failure.csv};
        \addlegendentry{\glspl{ue} 1-10}
        \addplot [c7, very thick, mark=*, mark repeat=10,mark options={solid,scale=1}] table 
        [   x = delta_time,
            y = ues_1,
        ] {data/failure.csv};
        \addlegendentry{\glspl{ue} 11-20}
        \addplot[c4!50!black, ycomb,dashed,very thick,no marks] coordinates{(19.5, 61)};
        \addplot[c4!50!black, ycomb,dashed,very thick,no marks] coordinates{(34.5, 61)};
        \addplot[c4!50!black, ycomb,dashed,very thick,no marks] coordinates{(60, 61)};
        
    \end{axis}

\end{tikzpicture}
    \caption{RTT from the \gls{iab}-nodes and the \glspl{ue} before, during, and after the link failure. The vertical lines indicate the time of fault and the time of the two recoveries.}\label{fig:fail_measures}
\end{figure}

Figure \ref{fig:fail_measures} shows how the \gls{rtt} of the transmitted packets is affected during and after the reconfiguration of the network. We average the \gls{rtt} on a rolling window of 10 s and across \glspl{ue} for each node. 
At $t=0$ both \gls{iab}-nodes report a \gls{rtt} of roughly 10 ms and their \glspl{ue} measure an \gls{rtt} of roughly 24 ms, which accounts for one more wireless hop and some switching time. At time $t=19$ s, we induce the failure of the link between \gls{iab}-node 2 and \iabd 2, which interrupts the successful transmission of the \gls{icmp} packets and triggers the reconfiguration of the network from the rApp. The reconfiguration of the OAI stack takes roughly 15 s, albeit being fully automated. Around second $t=35$ s, \iabn 2 is able to reach the core again, with an average \gls{rtt} of 22 ms, due to the additional hop. At that point, also the software-defined \glspl{ue} stack restarts, with their \gls{rtt} increased to an average of 47 ms, corresponding to 3 hops of distance to the core and two switching delays.

Despite the current state of OAI affects the performance of the reconfiguration, this experiment fully confirms the viability of our approach in a real-world network scenario based on the O-\gls{ran} specifications.

\section{Conclusions}
\label{sec:conclusions}

Next-generation wireless networks require optimized and intelligent management for their complex backhaul configurations, mixing wired and wireless components.  
This paper proposed a novel toolchain to plan a mobile backhaul, based on open data to model 3D scenarios, optimization strategies to plan the network, and an O-RAN rApp for dynamic network control. We have shown that the optimization problems can be solved for realistic networks of density up to 60 \glspl{gnb}/km$^2$, and this can lead to the reduction of a substantial number of fiber drops. The models are fully configurable in terms of depth of the \gls{iab} tree and required robustness, and can be used in greenfield or brownfield scenarios, and the approach has been prototyped in a realistic O-RAN-based network environment.

\bibliographystyle{IEEEtran}
\bibliography{bibliography.bib}

\end{document}